\documentclass[%
 reprint,
superscriptaddress,
 amsmath,amssymb,
 aps,
]{revtex4-1}
\usepackage{floatflt}
\usepackage{subfigure}
\usepackage{pdfpages}
\usepackage[section]{placeins}
\usepackage{graphicx}
\usepackage{dcolumn}
\usepackage{bm}
\usepackage{verbatim}
\usepackage{mathtools}
\usepackage{multirow}
\usepackage{bigdelim} 

\begin{document}

\preprint{APS/123-QED}

\title{Surface Conductivity of Si(100) and Ge(100) Surfaces Determined from Four-Point Transport Measurements Using an Analytical N-Layer Conductance Model}

\author{Sven Just}
\affiliation{Peter Gr\"{u}nberg Institut (PGI-3) and
JARA-Fundamentals of Future Information Technology,
Forschungszentrum J\"{u}lich, 52425 J\"{u}lich, Germany}

\author{Helmut Soltner}
\affiliation{Central Institute of Engineering, Electronics and Analytics (ZEA-1), 
Forschungszentrum J\"{u}lich, 52425 J\"{u}lich, Germany}

\author{Stefan Korte}
\affiliation{Peter Gr\"{u}nberg Institut (PGI-3) and
JARA-Fundamentals of Future Information Technology,
Forschungszentrum J\"{u}lich, 52425 J\"{u}lich, Germany}

\author{Vasily Cherepanov}
\affiliation{Peter Gr\"{u}nberg Institut (PGI-3) and
JARA-Fundamentals of Future Information Technology,
Forschungszentrum J\"{u}lich, 52425 J\"{u}lich, Germany}

\author{Bert Voigtl\"{a}nder}
\email[Corresponding author: ]{b.voigtlaender@fz-juelich.de}
\affiliation{Peter Gr\"{u}nberg Institut (PGI-3) and
JARA-Fundamentals of Future Information Technology,
Forschungszentrum J\"{u}lich, 52425 J\"{u}lich, Germany}




\date{\today}

\begin{abstract}
	An analytical N-layer model for charge transport close to a surface is derived from the solution of Poisson's equation and used to describe distance-dependent electrical four-point measurements on the microscale. As the N-layer model comprises a surface channel, multiple intermediate layers and a semi-infinite bulk, it can be applied to semiconductors in combination with a calculation of the near-surface band-bending to model very precisely the measured four-point resistance on the surface of a specific sample and to extract a value for the surface conductivity. For describing four-point measurements on sample geometries with mixed 2D-3D conduction channels often a very simple parallel-circuit model has so far been used in the literature, but the application of this model is limited, as there are already significant deviations, when it is compared to the lowest possible case of the N-layer model, i.e. the 3-layer model. Furthermore, the N-layer model is applied to published distance-dependent four-point resistance measurements obtained with a multi-tip scanning tunneling microscope (STM) on Germanium(100) and Silicon(100) with different bulk doping concentrations resulting in the determination of values for the surface conductivities of these materials. 

\end{abstract}

\pacs{Valid PACS appear here}
\maketitle


\section{Introduction}

Due to the downscaling of modern nanoelectronic devices the surface-to-volume ratio increases continuously and the surface becomes increasingly important as an additional conductance channel for charge transport. To assess the influence of this surface channel on the device performance or even be able to use it as a functional unit, a reliable value for the two-dimensional surface conductivity has to be known. However, the determination of the surface conductivity from electrical four-point measurements is quite a challenging task, as the main difficulty is to separate the 2D conductance at the surface from the conductance through other channels, e.g. the bulk and the space charge layer. 

Often indirect measurement methods are used for the separation of the 2D conductance at the surface, but these methods have special requirements on the material and the preparation of the sample under study. For example, one method for separating the surface conductivity is based on the comparison of measurements before and after quenching the surface states by adsorption of atoms or molecules \cite{HasegawaA,HasegawaB,Petersen,Hasegawa1,Wolkow}. The adsorption species has to be chosen specifically for the material under study and for the quenched system several conditions have to be carefully confirmed. First, all of the surface states have to be quenched and, secondly, the conductivity of the near-surface space charge region has to remain unchanged under the influence of the adsorbed surface layer. Thirdly, no additional surface conductance has to be induced by the adsorbed layer. 
If one of these conditions is not fulfilled, the experiments based on the difference method can result in underestimated values for the surface conductivity.  

Here, we present a generic N-layer conductance model, free of such requirements, for describing the measured four-point resistance on samples consisting of a surface channel, a space charge region due to the near-surface band-bending and a semi-infinite bulk. 
No special sample preparation is necessary and the model can directly be applied to the raw data, which in combination with a calculation of the conductivity profile in the space charge region permits to extract the value for the surface conductivity from distance-dependent four-point measurements. 

First, we compare a very simple model often used to describe measurements on samples with mixed 2D-3D conduction channels, the parallel-circuit model, to the N-layer model, and point out that the application of the parallel-circuit model is very limited, as there are already significant deviations if the N-layer model is reduced to the simplest case of a 3-layer model ($N=3$).
Secondly, we apply the N-layer model to different distance-dependent four-point measurements from the literature obtained with a multi-tip scanning tunneling microscope on the semiconductors Ge(100) and Si(100) with different types and concentrations of doping, and determine values for the surface conductivity of these materials. 
The analytical derivation of the N-layer model is shown in detail in the appendix. 

\section{Mixed 2D-3D conduction channels}

\begin{figure}[t!]
\centering
\includegraphics[width=0.435\textwidth]{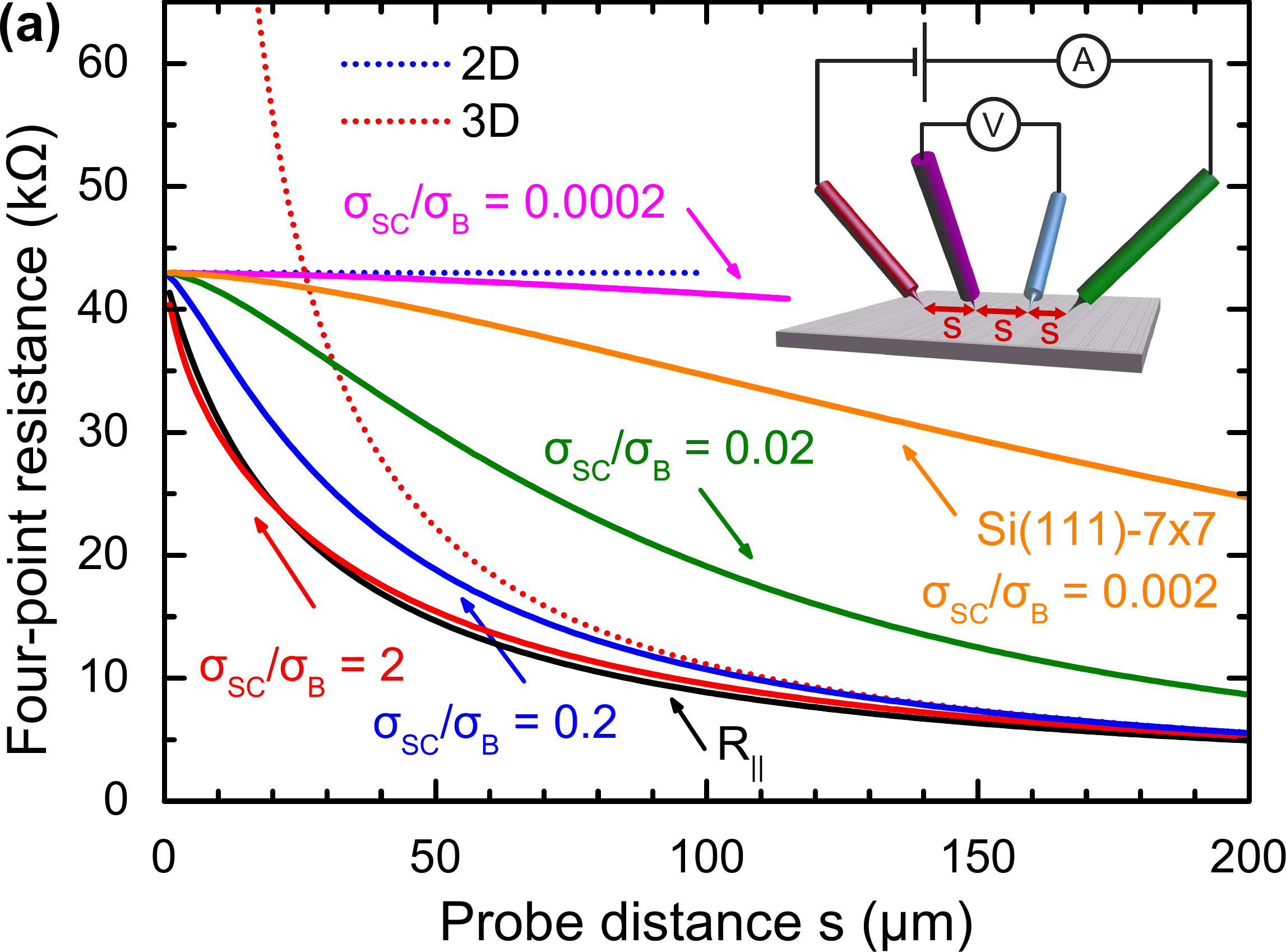}
\includegraphics[width=0.445\textwidth]{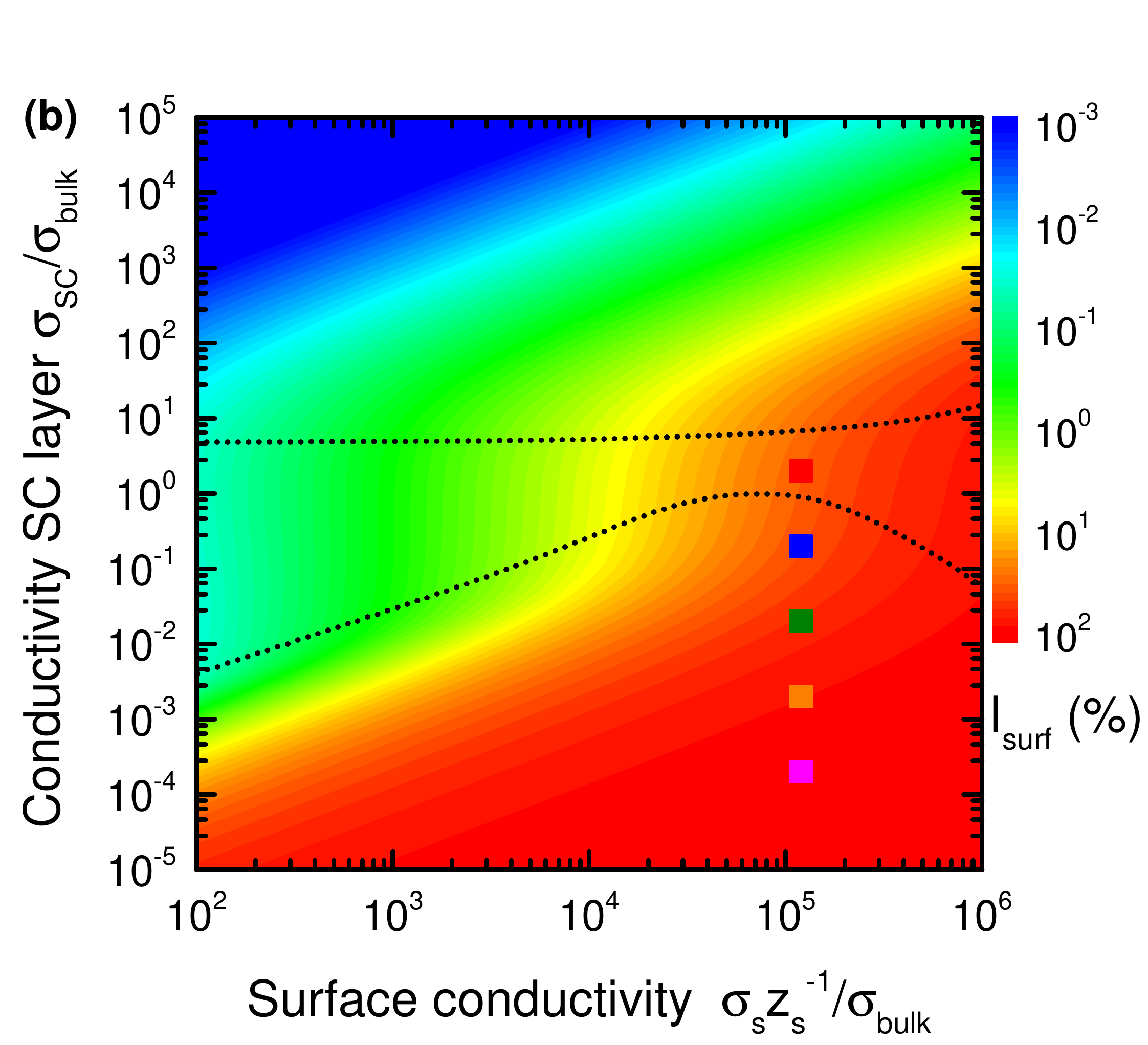}
\caption{(Color online) (a) Calculated four-point resistance of the Si(111)-($7\times7$) surface with a bulk conductivity of $\sigma_B = 0.14\,\mathrm{S/m}$ and a surface conductivity of $\sigma_{S} = 5.14 \cdot 10^{-6}\,\mathrm{S/\square}$ as a function of the equidistant probe distance $s$ and with the ratio $\sigma_{SC}/\sigma_B$ between the conductivities of the space charge layer and the bulk as additional parameter (colored curves). The orange curve located between the two limiting cases of pure 2D and pure 3D conductance (dotted blue and red curves) is based on measurements \cite{Just}, while the magenta, green, blue and red curves correspond to variations of the ratio $\sigma_{SC}/\sigma_B$ over several orders of magnitude. The black curve results from the description by the parallel-circuit model without considering an additional space charge layer between surface and bulk. In the inset, the equidistant linear tip arrangement with the outer current-injecting tips and the inner voltage-measuring tips is shown.    
(b) Calculated percentage of surface current $I_{\mathrm{surf}}$ as function of the ratios $\sigma_S\,z_S^{-1}/\sigma_B$ between the surface conductivity and the bulk ($z_S = 3\,\mathrm{\AA}$), and $\sigma_{SC}/\sigma_B$ between the conductivity of the space charge layer and the bulk. The colored points correspond to the position of the curves in (a). Inside the region marked by the two dotted lines the parallel-circuit model can be applied for describing the four-point resistance on the surface with an error of less than 10\%. }
\label{fig1}
\vspace{-2ex}
\end{figure}

\begin{figure*}[t!]
\centering
\includegraphics[height=0.345\textwidth]{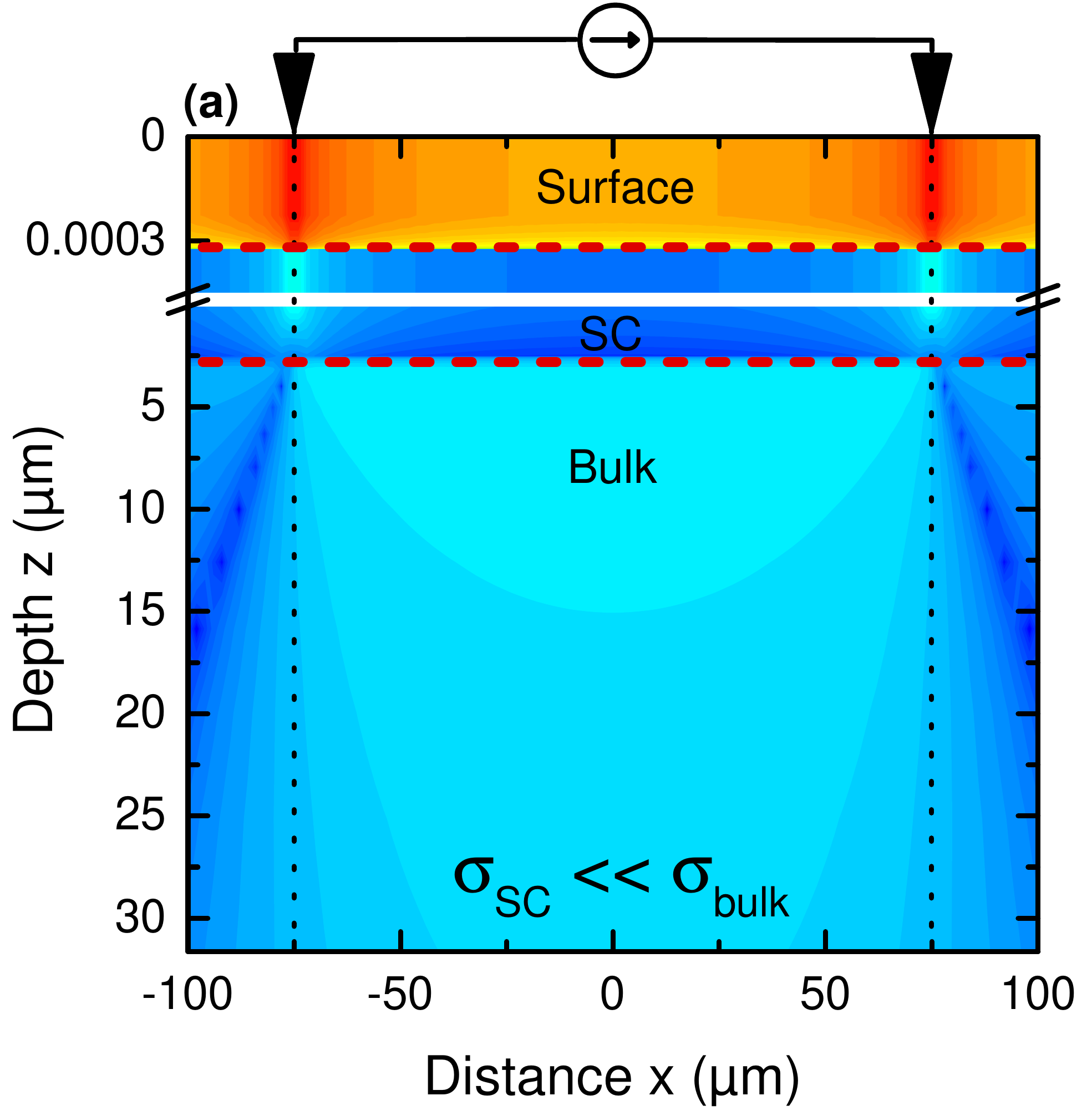}
\includegraphics[height=0.345\textwidth]{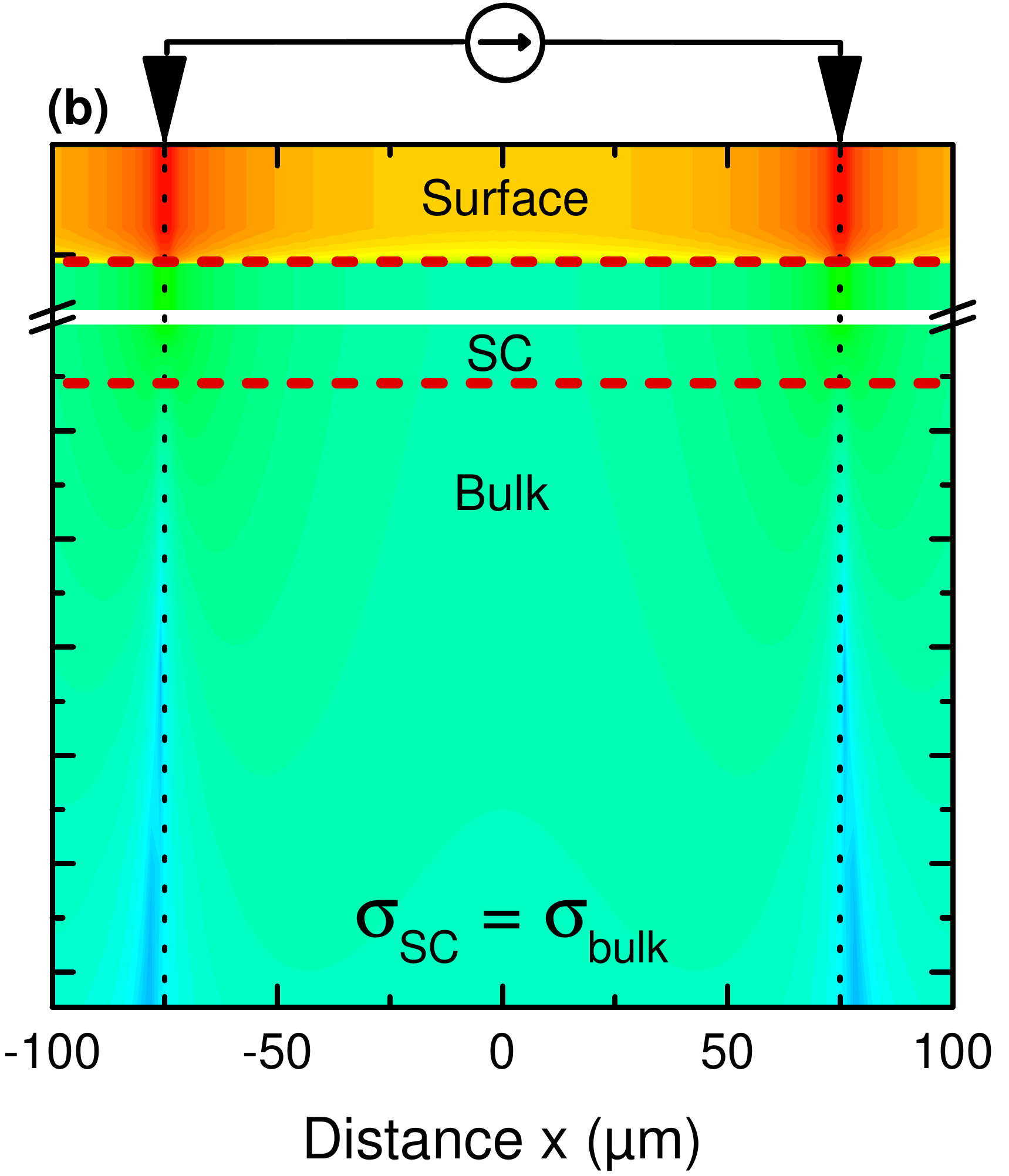}
\includegraphics[height=0.345\textwidth]{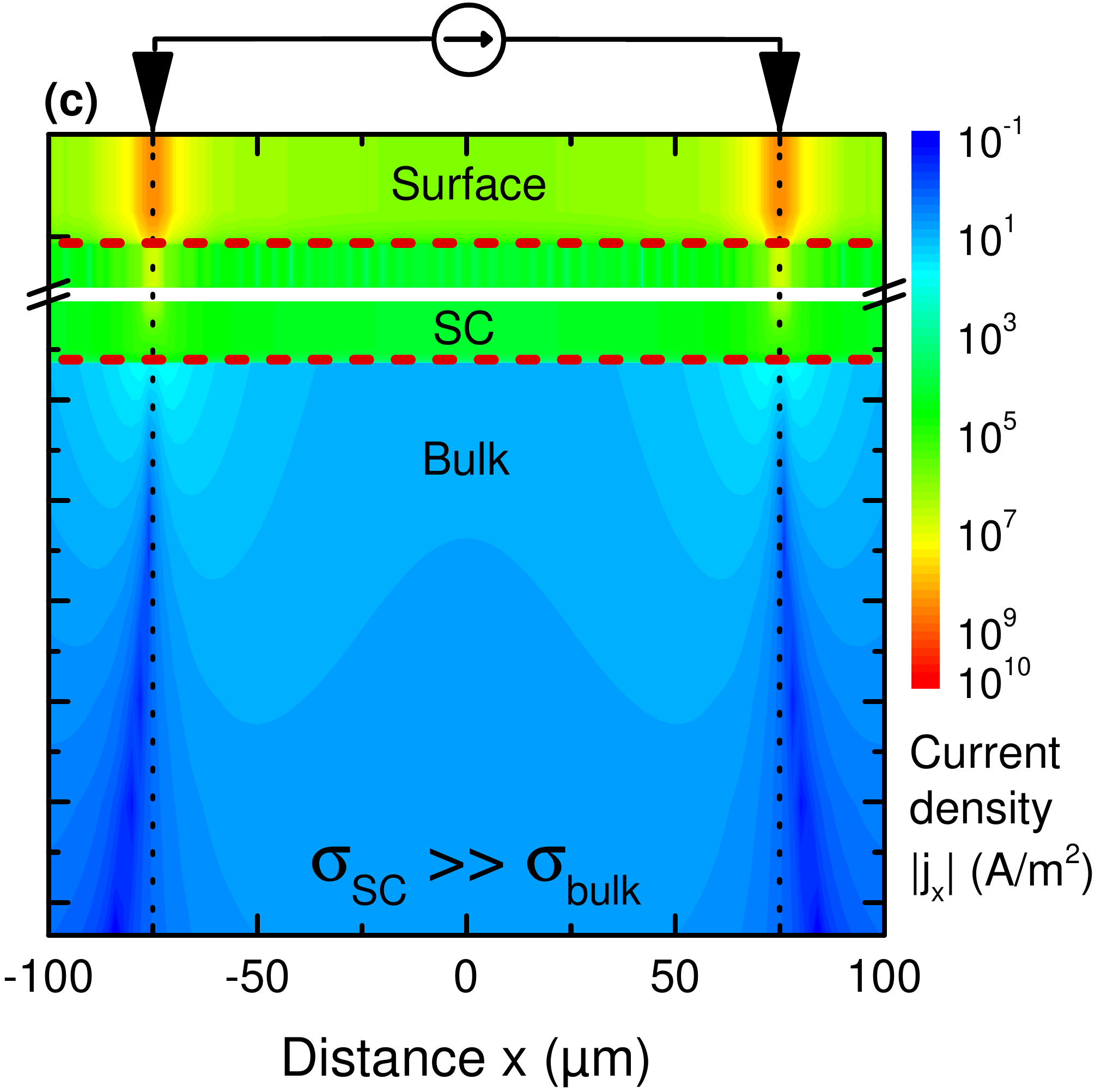}
\caption{(Color online) Color plots of the absolute value of the in-line component of the current density $\mathbf{j}(x,y,z)$ in the xz-plane as a function of depth $z$ into the sample and lateral distance $x$ along the tip positioning line. The current density is calculated from the 3-layer model for a distance $3s = 150\,\mu\mathrm{m}$ of the current-injecting tips, and for a sample with a bulk conductivity $\sigma_B = 0.14\,\mathrm{S/m}$, a surface conductivity $\sigma_{S} = 5.14 \cdot 10^{-6}\,\mathrm{S/\square}$ and an average thickness $z_2 = 2.5 \,\mathrm{\mu m}$ of the intermediate space charge layer. The average conductivity of the intermediate space charge layer is varied in the three cases (a) - (c) showing the significant influence of the space charge region on the vertical current distribution in the sample. According to the 3-layer model the red dashed lines indicate the interfaces between the surface, the space charge layer and the bulk. The black dotted vertical lines mark the position of the current-injecting tips on the surface. 
(a) In the case of a very low conducting space charge layer with $\sigma_{SC} \ll \sigma_B$ ($\sigma_{SC} = 2.5 \cdot 10^{-4}\,\mathrm{S/m}$) the majority of the current flows through the surface even if the bulk is highly conductive, as the space charge region acts as a blockade for the injection into the bulk and an enhanced 2D transport can be observed.  
(b) If $\sigma_{SC} = \sigma_B$, there is effectively no space charge region and the current flow through the bulk takes place according to the bulk conductivity. In this case the four-point resistance on the surface can be approximated by the parallel-circuit model.    
(c) If the space charge layer is highly conductive with $\sigma_{SC} \gg \sigma_B$ ($\sigma_{SC} = 2.5 \cdot 10^{2}\,\mathrm{S/m}$), the current flows not only through the surface, but also equally through the space charge layer, while the current in the bulk is again reduced. }   
\label{fig3}
\end{figure*}

For pure 2D or pure 3D charge transport, there exist simple analytic relations between the measured four-point resistance and the conductivity. For an equidistant probe setup with a distance $s$ between the tips, the following equations are obtained for a 2D sheet and a 3D half-space \cite{Schroder}, respectively 
\begin{equation}\label{eq:1}
R^{4p}_{2D}=\frac{\ln 2}{\pi \sigma_{2D}},\quad \mathrm{and}\quad R^{4p}_{3D}=\frac{1}{2 \pi \sigma_{3D}} \cdot s^{-1} 
\end{equation}
with the 2D surface conductivity $\sigma_{2D}$ and the 3D bulk conductivity $\sigma_{3D}$. The equation for the 2D case shows a constant four-point resistance, independent of the probe spacing, while the conductance through a 3D channel depends on the spacing $s$. Due to this characteristic probe-spacing dependency, it is possible to distinguish between 2D and 3D channels from distance-dependent four-point measurements.    
However, if a sample consists of a mixed 2D-3D geometry, e.g. a conducting sheet on a conducting substrate, these two equations cannot be applied any more. Often, a simple approximation of a parallel-circuit consisting of the four-point resistance of the surface and the bulk according to Eq.(\ref{eq:1}) is used \cite{Perkins,Wells3} 
\begin{equation}\label{eq:2}
	R^{4p}_{\parallel}(s) =  \left(\frac{1}{R^{4p}_{2D}} + \frac{1}{R^{4p}_{3D}(s)}\right)^{-1}\,\mathrm{,}
\end{equation}
but this approach has restrictions and shortcomings, as it can be seen in the following. 

In the parallel-circuit model a complete separation of the surface conductance channel and the bulk is assumed. The splitting of the injection current between the surface and the bulk only takes place at the injection points and depends on the ratio of the four-point resistances of the two individual layers. However, the two-point resistance, and not the four-point resistance, should determine, which amount of current flows through the surface channel and which part through the bulk \cite{Polley}. Therefore, the exact current path through the sample depends also on details of the injection, e.g. the tip diameter, which are not included in the parallel-circuit model.
The most important point, however, is the fact that in the approximation of the parallel-circuit model the current is injected equally into the surface channel and the bulk, and any influence of a possible near-surface space charge region, which particularly exists in semiconductors, is neglected. But especially this space charge region has a significant influence on the charge transport through the sample, as it will be discussed in the following. 

A different approach presented in \cite{Durand} uses an approximation for the surface current to solve the current continuity equations for 2D and 3D resulting in a combination of both 2D and 3D conduction channels. This approach removes the artificial separation between surface and bulk and uses a real injection geometry with extended tips, but it takes only into account a two-layer structure consisting of the surface and the bulk, so that the results are very similar to the parallel-circuit model. Any additional conductivity distribution between the surface and the bulk caused by a space charge region is neglected, which is also the major restriction in the parallel-circuit model. For this reason, the model can only be applied, if no near-surface band-bending occurs and a sharp transition between surface and bulk exists.  

Another approach published in \cite{Szymonski2} attempts to describe the deviation from a pure 3D conductance behavior caused by an additional 2D channel with an expansion of distance-dependent terms, and introduces an effective conductivity consisting of the bulk conductivity and a value for the deviation from the pure 3D case. However, although this model may also be able to treat deviations caused by a near-surface space charge region, it is not suitable to determine a value for the surface conductivity, as the deviations from the pure 3D conductance are only indicated by one numerical value, which cannot be easily interpreted as a physical quantity. 

In \cite{Wells2} a computational method is described using no longer an analytical model for the four-point resistance but a finite element calculation for approximating the different conduction channels in the sample. In this case, also the near-surface space-charge layer between the surface channel and the 3D bulk can be taken into account. However, as the surface channel has only a depth of several \AA, while the space-charge layer may be extended up to several $\mu m$, very different length scales are involved, so that the finite element calculation of the complete sample geometry can be very sophisticated and computationally time consuming. 

The best way to point out the important role of the space charge region, which is especially important for semiconductors, and the limited applicability of a two-layer model, like the parallel-circuit model, is a comparison of the four-point resistance with the lowest N-layer model including the influence of the space charge region, i.e. the 3-layer model. Apart from the surface layer and the bulk region this 3-layer model uses only one additional layer to approximate the space charge region, but despite this quite rough approximation it is able to describe four-point resistance measurement values much better than the parallel-circuit model and was successfully applied to determine the surface conductivity of the Si(111)-($7\times7)$ surface \cite{Just}. 

In Fig. \ref{fig1} (a) the calculated distance-dependent four-point resistance for the Si(111)-($7\times7)$ surface on an n-doped substrate ($700\,\mathrm{\Omega cm}$) is shown (orange line) located between the two limiting cases of pure surface conductance (dotted blue line) and pure bulk conductance (dotted red line). The calculation is based on the 3-layer model with parameters obtained in \cite{Just} and assumes an equidistant linear tip configuration with a tip spacing $s$. Using the same parameters for surface and bulk conductivity the four-point resistance expected from the parallel-circuit model according to Eq. \ref{eq:2} is plotted as solid black line, which exhibits a very strong deviation from the curve based on the 3-layer model. The major reason for this behavior is the absence of the additional space charge layer between surface and bulk in the parallel-circuit model. In the case of the Si(111)-($7\times7)$ surface on an n-doped Si substrate with $\sigma_B = 0.14\,\mathrm{S/m}$ the ratio between the average conductivity of the space charge region $\sigma_{SC}$ and the bulk can be estimated as $\sigma_{SC}/\sigma_B = 0.002$ \cite{Just}. For smaller values of this ratio, the deviation of the 3-layer model from the parallel-circuit model increases and the calculated four-point resistance approaches the 2D case (magenta curve). On the other hand, if the ratio becomes larger, the deviation between the two models decreases (green and blue curves). But only if the ratio $\sigma_{SC}/\sigma_B$ is close to 1 (red curve), the deviation between both models is so small, that the parallel-circuit model can be used as approximation without a large error. This error is smallest, if the near-surface space charge region vanishes completely, and in this case the parallel-circuit model is a suitable simple approach to approximate the four-point resistance of a two-layer structure consisting of a 2D and a 3D conduction channel.   

The significant influence of the space charge region can also be deduced from the amount of current flowing through the surface compared to the totally injected current. In Fig. \ref{fig1} (b) the calculated percentage of surface current is shown in dependence of the conductivity ratios between space charge layer and bulk $\sigma_{SC}/\sigma_B$ and the surface and bulk $\sigma_S\,z_S^{-1}/\sigma_B$ (thickness of surface layer $z_S \approx 3\,\mathrm{\AA}$) for a constant tip distance of $s = 50\,\mu\mathrm{m}$. The calculation is again based on the 3-layer model and on parameters obtained in \cite{Just} for the measurements of the Si(111)-($7\times7$) surface.  
For a vanishing space charge layer, i.e. $\sigma_{SC}/\sigma_B \approx 1$, the amount of surface current approximately only depends on the ratio $\sigma_S\,z_S^{-1}/\sigma_B$ and increases with an increasing ratio. However, if the influence of the space charge layer becomes larger, i.e. if the ratio $\sigma_{SC}/\sigma_B$ deviates from 1, the contour lines in the plot get distorted, so that for large ratios the amount of surface current is reduced and for small ratios enhanced. 

The reason for this behavior is that the conductivity of the space charge layer controls the current injection into the bulk below.  
If the near-surface band-bending leads to a depletion zone or an inversion zone so that the average conductivity in the space charge region is significantly reduced compared to the bulk, then this region behaves as a blocking region preventing the injected current to flow through the bulk, even if it has a very high conductivity. This results in an enhanced surface domination of charge transport, which cannot be considered in the parallel-circuit model. 

In Fig. \ref{fig3} this behavior is visualized by the depth-dependent current density inside the sample. The absolute value of the in-line component of the current density $\mathbf{j}(x,y,z)$ in the xz-plane is plotted as function of depth $z$ into the sample and lateral distance $x$ along the tip positioning line. The calculation is based on the 3-layer model with the same parameters as used in Fig. \ref{fig1}(b) and a distance of $3s = 150\,\mu\mathrm{m}$ for the current injecting tips. For the first case in Fig. \ref{fig3}(a), a very low conducting space charge layer with $\sigma_{SC} \ll \sigma_B$ (thickness $z_{SC} = 2.5\, \mathrm{\mu m}$) is used for the calculation, and the result shows that the majority of the current flows through the surface layer (thickness $z_S = 3\, \mathrm{\AA}$), whereas only a very small amount of current is injected through the space charge layer into the bulk. The current density inside the bulk material is one order of magnitude lower than in the case of a vanishing near-surface band-bending, where the space charge layer coincides with the bulk ($\sigma_{SC} \approx \sigma_B$), which is depicted in Fig. \ref{fig3} (b). 

On the other hand, if an accumulation zone is formed near the surface with a high conductivity compared to the bulk, this region can act as an additional conductance channel totally surpassing the current flow through the  bulk and also reducing the current through the surface states. In this case shown in Fig. \ref{fig3} (c), where $\sigma_{SC} \gg \sigma_B$, the current flow through the bulk is again reduced by an order of magnitude, while not only transport through the surface states but also through the space charge region is now preferred equally. As the space charge layer has a finite thickness, the current transport may seem to be purely 2-dimensional for larger probe spacings and the usage of the parallel-circuit model for the four-point resistance on such a system would result in a largely overestimated value for the surface conductivity. 

In conclusion, the parallel-circuit model has only a very limited applicability within a certain range of conductivity parameters, where the space charge region does not play a significant role for the current transport. In Fig. \ref{fig1}(b) the dotted lines indicate the region, inside which the parallel-circuit model can be applied to four-point resistance measurements with an error of less than 10\%. Inside this region, the contour lines of the color plot are approximately perpendicular to the x-axis indicating that the surface current is nearly independent of the ratio $\sigma_{SC}/\sigma_B$, which is an essential requirement for the application of the parallel-circuit model. For comparison, the four colored points indicate the positions of the resistance curves from Fig. \ref{fig1} (a). Only the red curve, which is very close to the parallel-circuit model, is located inside the dotted region, while the orange curve representing a measurement of the Si(111)-($7\times7$) surface on an n-doped substrate is clearly outside the region. 

Although the 3-layer model is obviously better suitable to describe measurement data over a wider range of conductivity parameters than the parallel-circuit model, it still has a basic restriction: the very rough description of the space charge region by only a single layer. Especially for semiconductors, which can have a very strong band-bending near the surface, this can be a major drawback. For this reason, the 3-layer model should be refined by introducing more layers resulting in an N-layer model, which is discussed in the following section.  

\section{The N-layer model} 

The 3-layer model offers only a rough approximation of the space charge region described by only a single layer with an average conductivity and average thickness. However, the conductivity profile in this region can exhibit a very strong dependence on the z-position, and, especially, if an inversion layer is formed in the near-surface region, the description by a single layer is not sufficient any more. Therefore, we try to approximate the space charge region by more than one layer and present an N-layer model for charge transport consisting of a thin surface layer, $N-2$ layers for the near-surface space charge region, and a semi-infinite bulk. Such a multi-layer model was first proposed by Schumann and Gardner \cite{Schumann, Schumann2, Gardner} and primarily applied to the method of spreading resistance measurements \cite{Leong, Berkowitz, Vandervorst}, but also extended to four-point measurements \cite{Wang} for determining individual sheet conductivities. However, as far as we know, it has not yet been used for obtaining the conductivity of surface states of semiconductors in combination with a calculated conductivity profile of the space charge region as input. 

A detailed description and mathematical derivation of the N-layer model is shown in the appendix \ref{appendix}. 
In the following section, the application of the N-layer model is demonstrated and it is used to obtain values for the surface conductivity of the Ge(100) and Si(100) surfaces.

\section{Application of the N-layer model}

The advantage of the N-layer model is that it can be used for evaluation of all distance-dependent four-probe resistance measurements without the need of any special sample preparation before the measurement, e.g. in order to quench the surface states \cite{Hasegawa1,HasegawaA,HasegawaB,Petersen}, or special measurement conditions, e.g. varying the temperature \cite{WellsPRL,Tanikawa,Yoo}. For this reason, we apply the N-layer model to already published data of the semiconductor surfaces Ge(100) and Si(100), which were described previously by either pure 2D or pure 3D conductance, but not by a mixed transport channel. In combination with the N-layer model, it is now possible to take into account simultaneously the current transport through the 2D surface and through the 3D bulk both influenced by the presence of the near-surface space charge layer, and to determine values for the surface conductivities of the materials from these measurements. 

\subsection{Germanium(100)}

\begin{figure}[t!]
\centering
\includegraphics[width=0.44\textwidth]{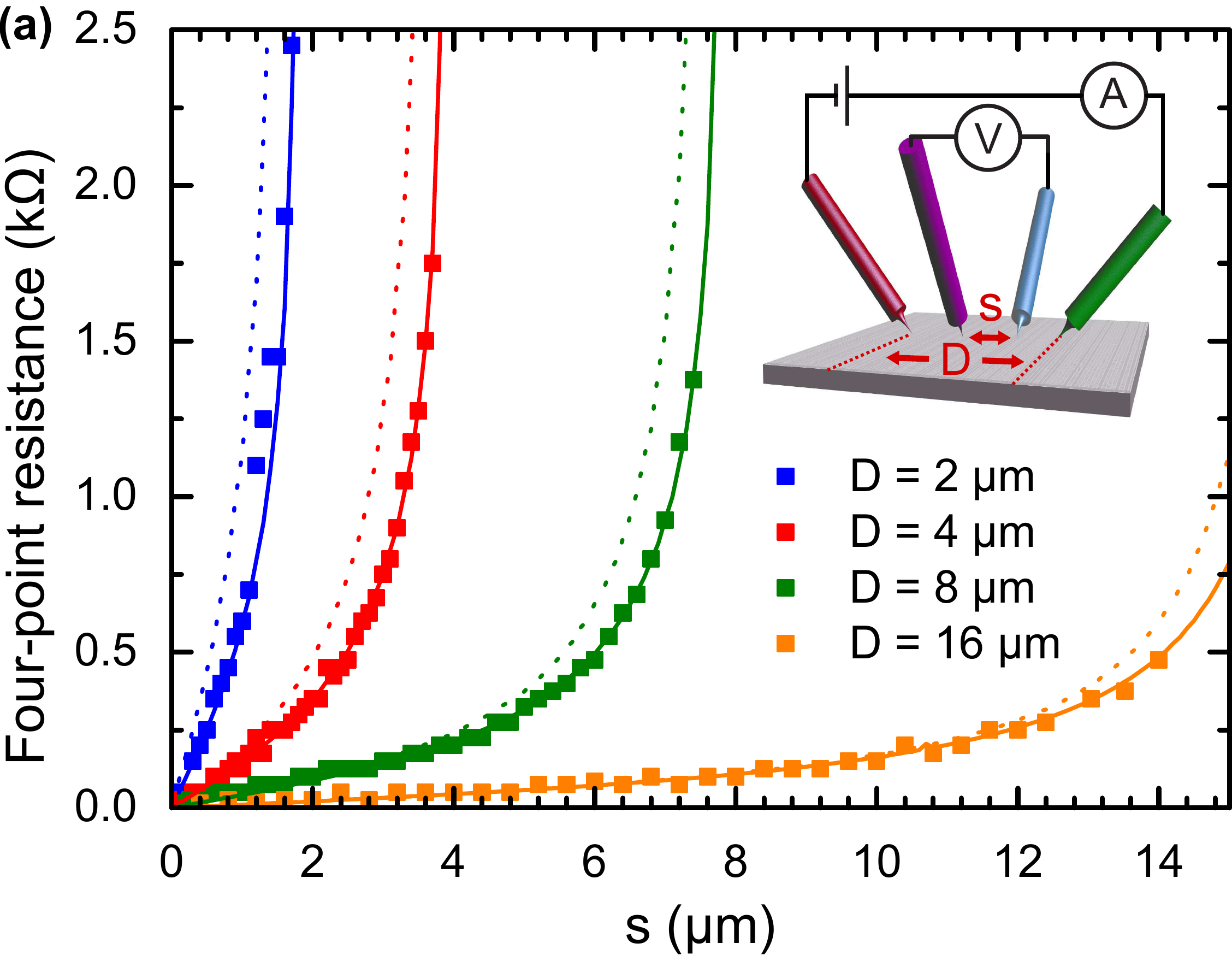}
\vspace{1ex}
\includegraphics[width=0.47\textwidth]{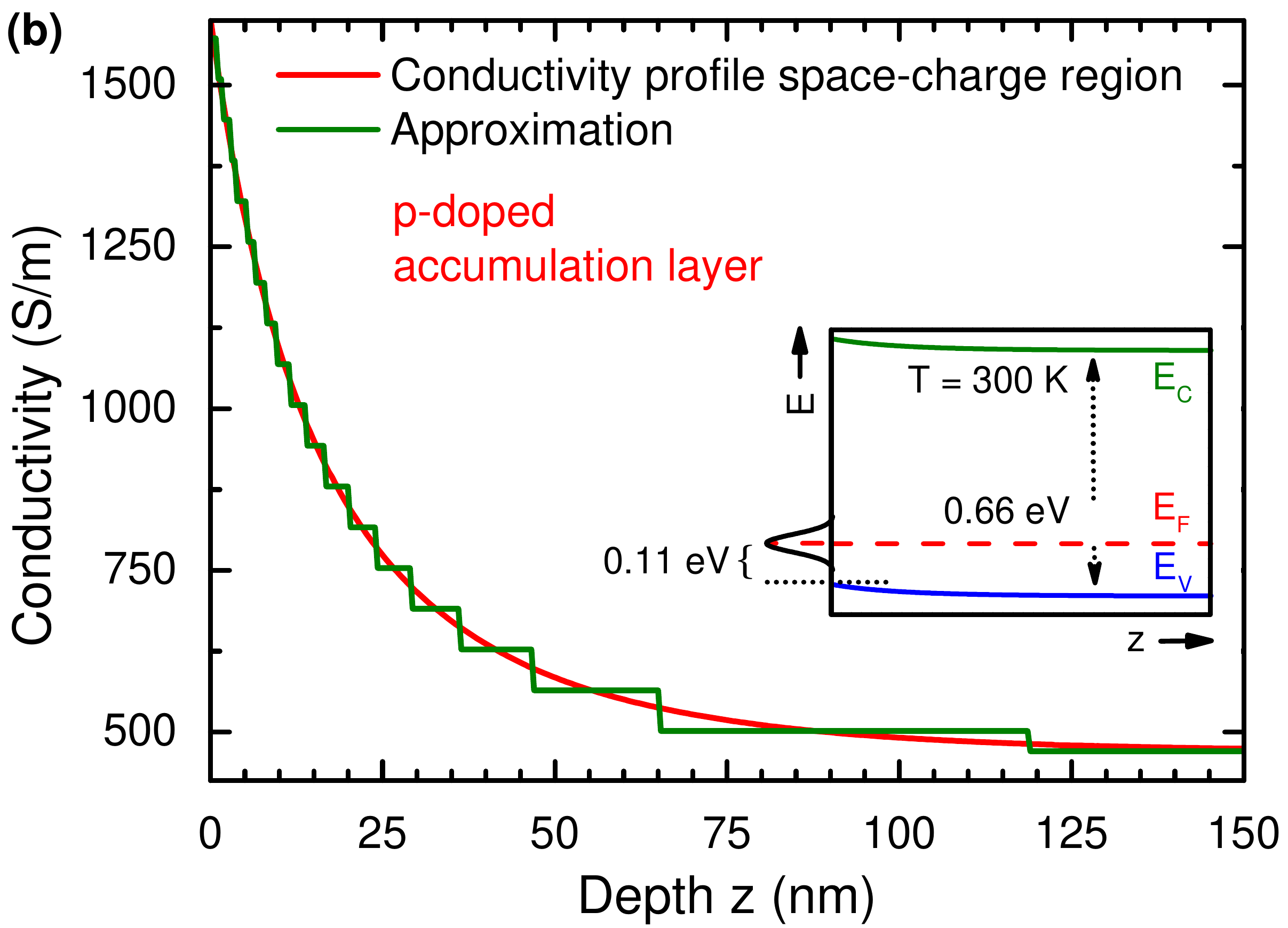}
\caption{(Color online) (a) Four-point resistance of a p-doped Ge(100) sample (nominal bulk resistivity $(0.1 - 0.5) \,\Omega\mathrm{cm}$) as function of probe distance $s$ between the inner voltage-measuring tips \cite{Wojtaszek}. Different colored data points correspond to different distances $D$ in the symmetric linear tip configuration shown in the inset. The solid lines represent one single fit to all data points using the N-layer model for charge transport, which results in a value for the surface conductivity of $\sigma_{S} = (2.9 \,\pm\,0.6) \cdot 10^{-4}\,\mathrm{S/\square}$ and for the bulk resistivity of $\rho_B = (0.22\,\pm\,0.01) \, \mathrm{\Omega cm}$. The dotted lines indicate the expected four-point resistances for a vanishing surface conductance channel, i.e. $\sigma_S = 0 $, taking into account only the space charge region and the bulk. (b) The calculated conductivity profile of the space charge layer as function of the depth $z$ into the sample starting from the surface. This profile is approximated with $N = 20$ layers and used as input for the N-layer model. The band diagram in the inset shows the surface pinning of the Fermi level $E_F$ (red) located $0.11\,\mathrm{eV}$ above the valence band edge and the resulting near-surface band-bending of the conduction band $E_C$ (green) and the valence band $E_V$ (blue).}  
\label{fig5} 
\end{figure}

Distance-dependent four-point transport measurements on the Ge(100) surface were published by Woj\-taszek \textit{et al.} \cite{Wojtaszek}. They used a room-temperature, ultra-high vacuum multi-tip STM and carried out four-point resistance measurements on Ge(100) substrates with different bulk doping concentration and type. A symmetric linear probe configuration was used, where  the outer current-injecting tips have a distance $D$ and the inner voltage-measuring tips are separated by the distance $s$. The complete setup is symmetric with respect to the centre plane of the tip positioning line. 
In Fig. \ref{fig5}(a), the experimental data for a p-type Ga-doped sample with a nominal bulk resistivity of $0.1 - 0.5\, \Omega\mathrm{cm}$ are shown \cite{Wojtaszek}. The measured four-point resistance is plotted as a function of the spacing $s$ between the voltage-measuring tips and with the distance $D$ between the current-injecting tips as additional parameter. In the framework of the publication \cite{Wojtaszek}, these data were described by a pure 3D conductance channel. However, it was mentioned that there were some systematic deviations from the 3D model, which increasingly appear, if the voltage-measuring tips approach the positions of the current-injecting tips, i.e. $s/D \ge 0.7$, but the origin of these deviations could not be explained quantitatively. 
In fact, for the symmetric linear tip configuration, it is particularly the region with a ratio $s/D$ close to 1, where the setup is most sensitive to surface transport and a possible surface conductance channel would have the most influence on the measured four-point resistance. So, it is reasonable to assume that the observed deviations are caused by an additional 2D conductance channel through the surface states of the Ge(100)-(2$\times$1) surface, which cannot be considered by the pure 3D model.

\begin{figure}[t!]
\centering
\includegraphics[width=0.45\textwidth]{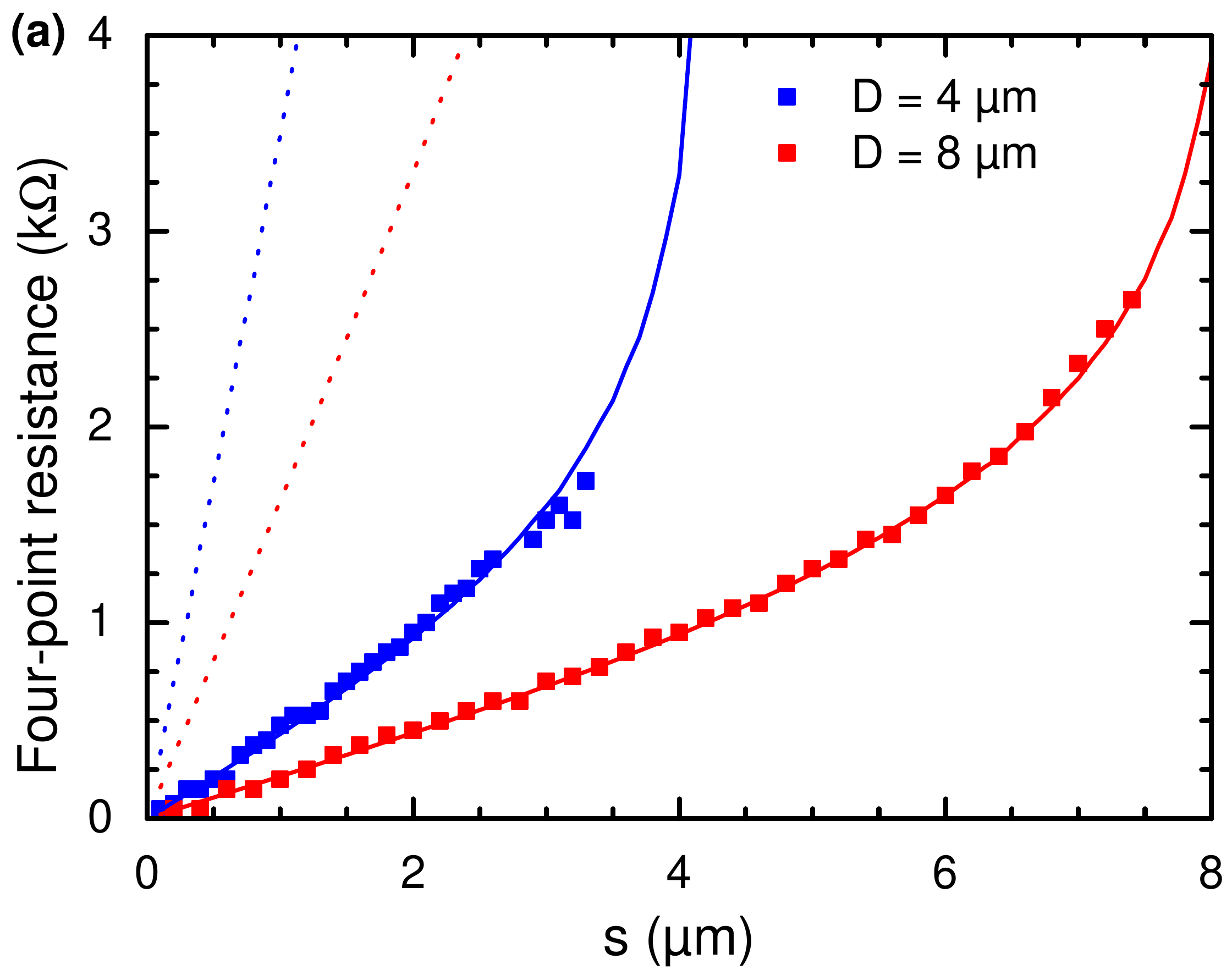}
\includegraphics[width=0.47\textwidth]{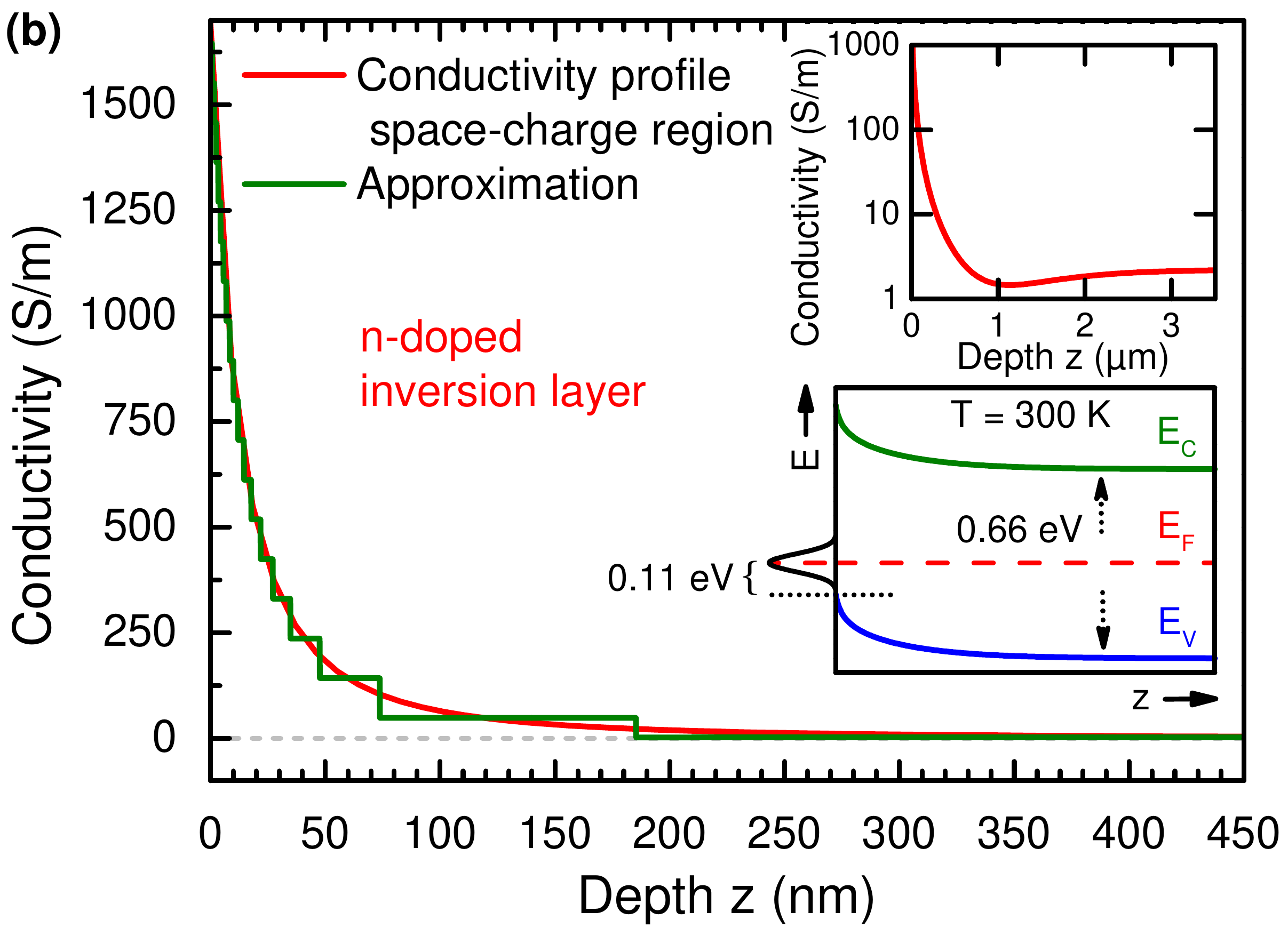}
\caption{(Color online) (a) Four-point resistance of an n-type doped, almost intrinsic Ge(100) sample (nominal bulk resistivity $\sim 45\,\Omega\mathrm{cm}$) as function of probe distance s between the inner voltage-measuring tips \cite{Wojtaszek}. Different colored data points correspond to different distances $D$ in the symmetric linear tip configuration (inset in Fig. \ref{fig5} (a)). The solid lines represent a single fit to all data points using the N-layer model for charge transport ($N = 20$), which results in a value for the surface conductivity of $\sigma_{S} = (3.4\,\pm\,0.2) \cdot 10^{-4}\,\mathrm{S/\square}$ and for the bulk resistivity of $\rho_B = (45\,\pm\,22)\,\mathrm{\Omega cm}$. The dotted lines correspond to the expected four-point resistances without any surface channel ($\sigma_S = 0$) taking into account only the bulk and the space charge region. (b) Calculated conductivity profile of the space charge region as function of the depth $z$ from the surface (red line). The approximated profile (green line) is used as input for the N-layer model. In the upper inset, the complete range of the conductivity profile of the space charge region exhibiting a shape of an inversion layer is shown. The lower inset depicts the surface pinning of the Fermi level $E_F$ (red) and the induced near-surface band-bending of the conduction band $E_C$ (green) and the valence band $E_V$ (blue).}
\label{fig6}
\end{figure}

In order to describe this additional 2D transport channel more quantitatively, we evaluate the existing data with the N-layer model. First, the near-surface band-bending of the p-type Ge(100) sample is calculated by solving Poisson's equation and using a Fermi level pinning at the surface of $\sim 0.11\,\mathrm{eV}$ above the valence band \cite{Tsipas,Tsipas2,Broqvist}. Fig \ref{fig5} (b) shows the resulting depth-dependent conductivity profile of the space charge region consisting of a near-surface accumulation layer. This conductivity profile is approximated by a step function of $(N-2)$ steps ($N = 20$) determining the values for $\sigma_n$ and $z_n$ to be used as input for the N-layer model (details in the appendix). For the symmetric linear tip setup the four-point resistance according to the N-layer model can be expressed as function of $s$ and $D$ by the equation 
\begin{align}
	R^{4p}(s,D) = & \frac{2}{I} \int_0^\infty \left[ a_0(k) + a_1(k) \right] \cdot \left[J_0\left(k\cdot\frac{D-s}{2}\right) \right. \nonumber \\[1ex]
& \left. - J_0\left(k\cdot\frac{D+s}{2}\right)\right] \, \mathrm{d}k\, \mathrm{,}  \label{r4psymm}
\end{align}
which is fitted to the measurement data resulting in the colored solid curves shown in Fig. \ref{fig5} (a). All four curves for the different values for the distance $D$ correspond to only a single fit with the surface conductivity $\sigma_S$ and the bulk conductivity $\sigma_B$ confined close to the range of the nominal values as free parameters. As the conductivity profile of the space charge region also depends on the bulk conductivity, an iterative fitting process is applied, which includes the calculation of the space charge region and the fit to the data in each iteration. For values of $\sigma_S = (2.9\,\pm\,0.6) \cdot 10^{-4} \mathrm{S}/\square$ and $\sigma_B = (460\,\pm\,11) \mathrm{S}/\mathrm{m}$ the iterative process converges and the best fit is obtained describing the data very precisely throughout the complete measurement range without any systematic deviations. A further advantage is the resulting single value for each of the parameters $\sigma_S$ and $\sigma_B$, which is sufficient to describe precisely all four resistance curves for the different distances $D$. In the case of a pure 3D model, as it is used for the fitting process in \cite{Wojtaszek}, it is not possible to model all four data sets with only one value for the bulk conductivity $\sigma_B$. The 3D fit has to be applied separately to each curve resulting in different values for $\sigma_B$ spreading by a relative deviation of $\sim$ 25\%. However, the measured bulk conductivity should not change during the variation of the tip configuration by the distance $D$ on the same substrate.
%
%
This reveals that, even if the transport in the sample is mostly 3D dominated due to the highly conductive bulk and the weak accumulation zone near the surface, a description of the data by a pure 3D model is not sufficient and an additional 2D channel has to be taken into account.  

For validating the results for the additional surface conductance channel and ensuring that the observed amount of two-dimensional conductance is not merely caused by the near-surface accumulation layer, the dotted colored curves in Fig. \ref{fig5} (a) correspond to the expected four-point resistance for a vanishing surface channel. In these curves, only the bulk conductivity and the conductivity profile of the space charge region according to Fig. \ref{fig5} (b) are taken into account, while the value for the surface conductivity $\sigma_S$ is set to zero. The clearly visible deviation of the dotted curves from the measurement data verifies that an additional 2D surface conductance channel is necessary for describing the measured four-point resistance, and, therefore, proves the existence of conducting surface states.  

Fig. \ref{fig6} (a) shows similar distance-dependent four-point resistance measurements on an n-type doped, almost intrinsic Ge(100) sample with a nominal bulk resistivity of $\sim 45\,\Omega\mathrm{cm}$ \cite{Wojtaszek}. As the measurement data show an enhanced two-dimensional character of conductance, a pure 2D model was used in \cite{Wojtaszek}, which was justified by the presence of a near-surface inversion layer totally preventing the current to be injected into the bulk and acting as a 2D channel, which confines the current close to the surface. However, any possible presence of an additional 2D surface channel caused by surface states was neglected. In this case, a further disentanglement between the conductivity of the near-surface p-type part of the inversion layer and the surface conductivity would be required.

So, we try again to describe the measurement data with the N-layer model. The calculated conductivity profile of the space charge region shows the expected inversion layer depicted in Fig. \ref{fig6} (b). For the calculation, the transition region between p-type and n-type of conduction has not been taken into account and only the absolute value of the conductivity is considered, but, as the majority of the current flows through the near-surface p-type part of the inversion layer and through the surface channel, this approximation should be suitable in the present case. The conductivity profile is described by a step function (green line) and used in combination with the N-layer model for a fit to the data according to Eq. \ref{r4psymm}. In Fig. \ref{fig6} (a), the two solid curves result from a single fit with the parameters $\sigma_S = (3.4\,\pm\,0.2)\cdot 10^{-4}\,\mathrm{S}/\square$ and $\sigma_B = (2.2\,\pm\,1.1)\,\mathrm{S}/\mathrm{m}$ and describe the data very precisely. For verification, the dotted lines shown in Fig. \ref{fig6} (a) again represent the expected four-point resistance without any additional surface channel ($\sigma_S = 0$). The very strong deviation from the measurement data indicates clearly that the observed transport behavior cannot only be caused by the enhanced conductivity close to the surface due to the inversion layer, but that there has to be an additional surface conductance channel also on the n-type sample.  

If the results for the p-type and n-type Ge(100) samples are compared, the values for the obtained surface conductivity coincide within the error limits. This is expected, as the surface states should not be influenced by the doping type of the substrate. Thus, this is another confirmation that really the conductivity of the surface states was determined. By combining the results of the p- and n-type sample, a more precise value for the surface conductivity of the Ge(100)-(2$\times$1) surface of $\sigma_{S,Ge(100))} = (3.1\,\pm\,0.6) \cdot 10^{-4}\,\mathrm{S}/\square$ can be obtained.

\subsection{Silicon(100)}

\begin{figure}[t!]
\centering
\includegraphics[width=0.41\textwidth]{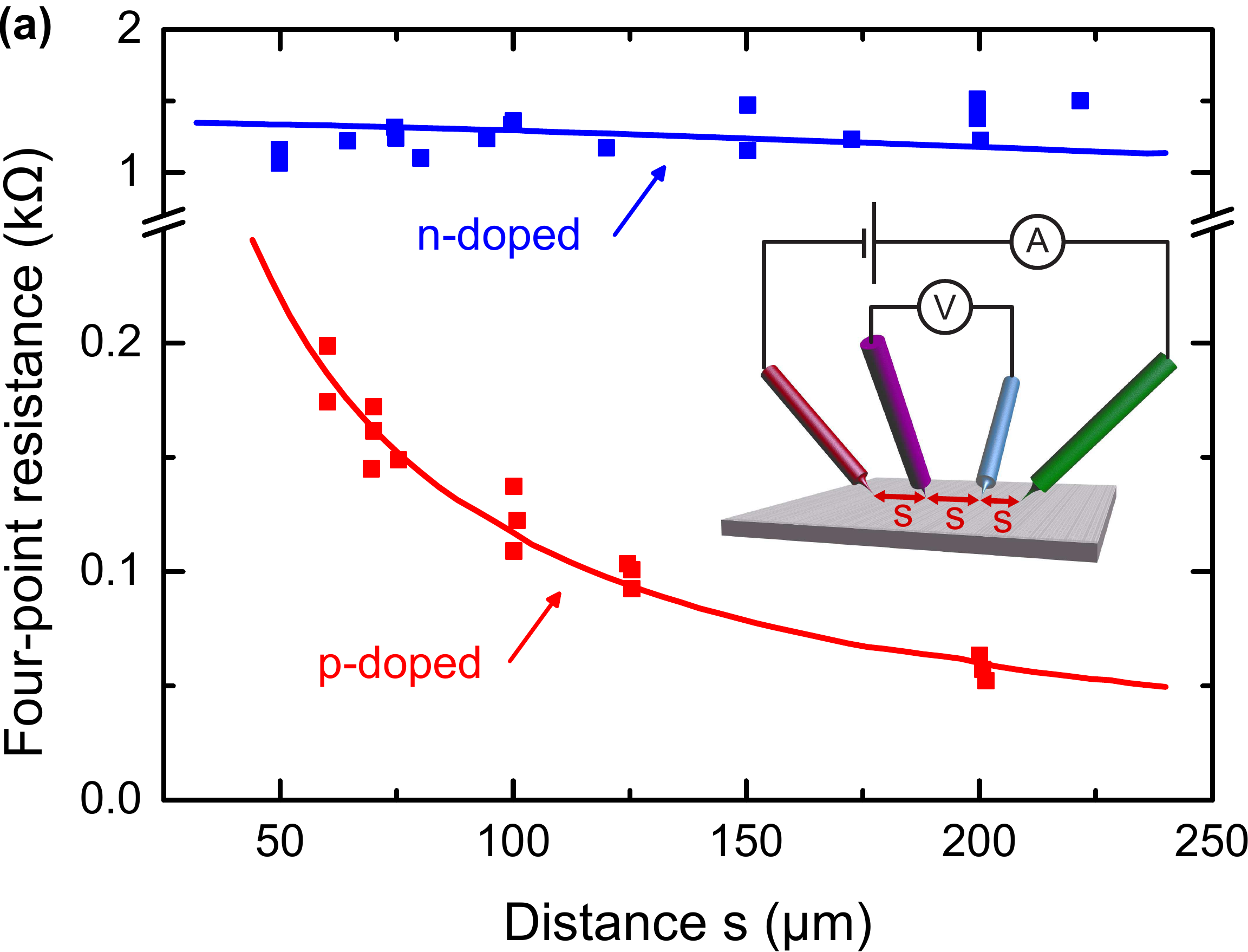}
\includegraphics[width=0.425\textwidth]{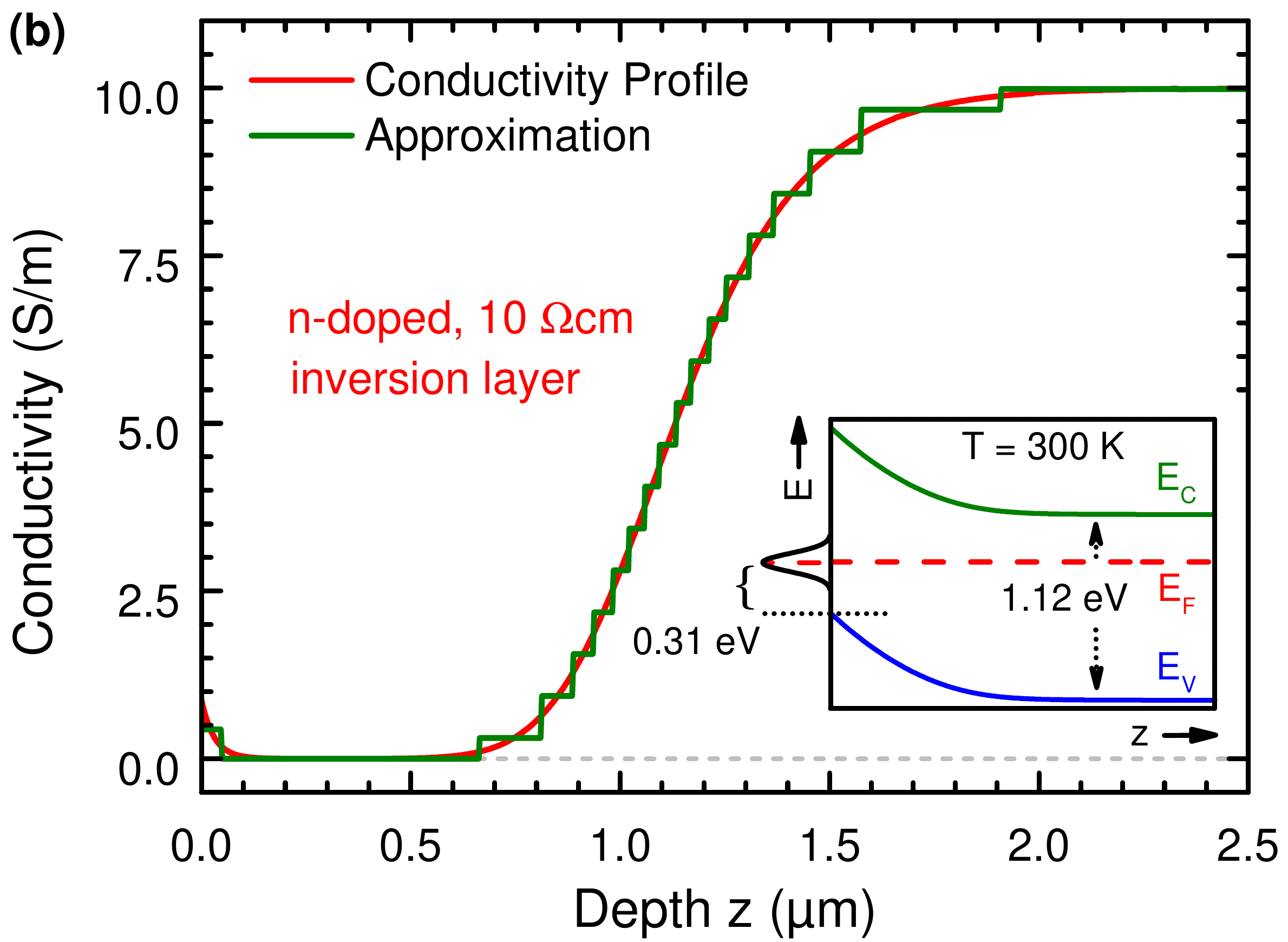}
\includegraphics[width=0.425\textwidth]{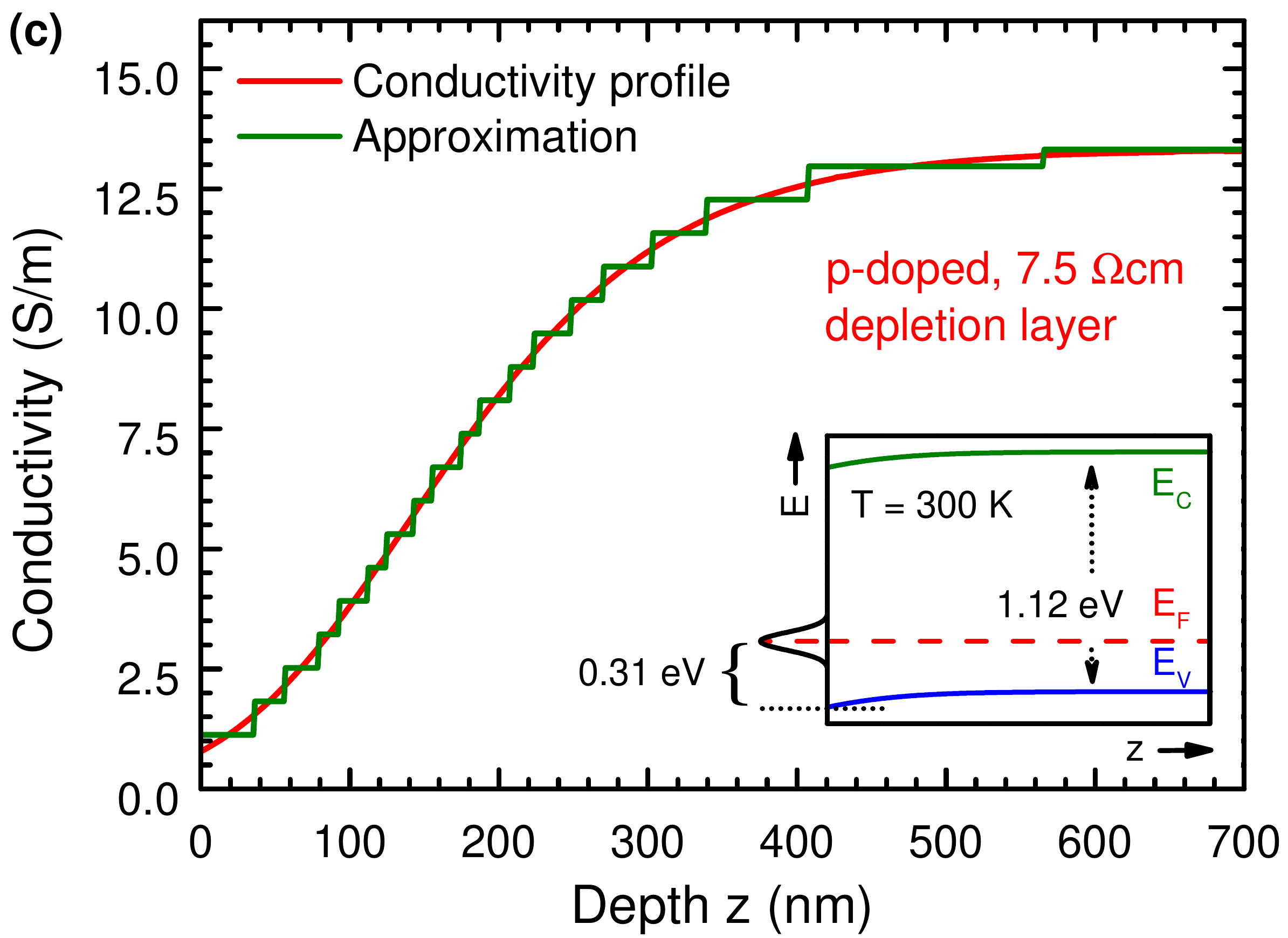}
\caption{(Color online) (a) Four-point resistance of a p-doped (red data points) and an n-doped (blue data points) Si(100)-(2$\times$1) sample (nominal bulk resistivity $(1-10)\,\mathrm{\Omega cm}$) as function of the equidistant probe distance $s$ reproduced from \cite{Polley}. Fits to the data (solid lines) based on the N-layer model result in a surface conductivity of $\sigma_{S} = (1.9\,\pm\,1.4) \cdot 10^{-4}\,\mathrm{S/\square}$ and in a bulk resistivity of $\rho_B = (7.5\,\pm\,0.9) \,\mathrm{\Omega cm}$ for the p-doped case, and in $\sigma_{S} = (1.6\,\pm\,0.4) \cdot 10^{-4}\,\mathrm{S/\square}$ and $\rho_B = (10\,\pm\,7.5)\,\mathrm{\Omega cm}$, respectively, for the n-doped sample. The inset shows the equidistant tip configuration. (b),(c) Calculated conductivity profiles of the space charge region for the p- and n-doped samples (red curves). The approximation by $N = 20$ layers (green curves) is used for the N-layer model. In the insets, the near-surface band-bending of the conduction band $E_C$ (green) and the valence band $E_V$ (blue) caused by the surface pinning of the Fermi level $E_F$ (red) due to the surface states located $\approx 0.31\, eV$ above the valence band edge is shown.}
\label{fig7}
\vspace{-0.7cm}
\end{figure}

Distance-dependent four-point resistance measurements on p-type and n-type doped Si(100) substrates were carried out by Polley \textit{et al.} \cite{Polley}. For the measurements, a room temperature, ultra-high vacuum multi-tip STM was used with a linear equidistant tip configuration with spacing $s$ between adjacent tips. The current was injected by the outer tips and the potential drop between the inner tips was measured. In Fig. \ref{fig7} (a), the measured four-point resistance is shown as a function of the tip distance $s$ for an n-type (blue points) and a p-type (red points) Si(100) substrate both with a nominal bulk resistivity of $(1-10)\,\Omega\mathrm{cm}$. Although the bulk doping concentrations of p- and n-type sample are similar, the observed transport behavior is completely different. In the p-type case, a 3D conduction channel is more dominant, while in the n-type case the majority of current flows through a 2D transport channel. Again, this was explained by the presence of an inversion layer in the n-type sample preventing the current to flow through the bulk. So, the measured data were described in \cite{Polley} by a pure 3D conductance model for the p-type substrate and by a pure 2D model in the n-type case. However, this approach cannot consider any possible mixed 2D-3D conductance channels through the space charge region and the bulk in both samples, and, especially, neglects the two-dimensional surface state, which should be present on the Si(100)-(2$\times$1) surface \cite{Martensson}.    

For refining the description of the measured data on the Si(100) substrates and for determining a value for the conductivity of the Si(100)-(2$\times$1) surface state, we use the N-layer model. Fig. \ref{fig7} (b) and (c) show the corresponding conductivity profiles of the space charge region for the p-type and n-type Si(100) substrates, respectively. For the calculation, a Fermi level pinning of the surface states of $\sim 0.31\,\mathrm{eV}$ above the valence band is used \cite{Yoo,Martensson,Himpsel2}. In the p-type case, a depletion zone is formed close to the surface, while in the n-type case an inversion layer separates the bulk from the near-surface region. Again, the pn-transition is not considered for the inversion layer, as the n-type bulk does not contribute significantly to current transport. The approximation of the conductivity profiles (green curves) is used as input for fitting the respective measurement data in Fig. \ref{fig7} (a) according to Eq. \ref{r4p-linear}. The results are depicted as solid curves in Fig. \ref{fig7} (a) and correspond to fitparameters for the surface conductivity of $\sigma_S = (1.9\,\pm\,1.4) \cdot 10^{-4}\,\mathrm{S}/\square$ and for the bulk conductivity, which is confined to the range of the nominal value, of $\sigma_B = (13.3\,\pm\,1.7) \,\mathrm{S}/\mathrm{m}$ for the p-type sample, and to values of $\sigma_S = (1.6\,\pm\,0.4) \cdot 10^{-4}\,\mathrm{S}/\square$ and $\sigma_B = (10\,\pm\,7.5) \,\mathrm{S}/\mathrm{m}$ in the n-type case. 
As the four-point resistance measurement for the p-type sample in the chosen tip distance range is not very surface sensitive, the determined value for the surface conductivity has quite a large error, even if the curve fits quite well to the data. The fitted curve for the n-type substrate shows some larger deviations due to a larger spread and a slight increasing behavior of the data, which might be caused by tip positioning errors or influence of the sample edges. However, the obtained value for the surface conductivity is more precise, as the transport behavior in the n-type sample is now more dominated by the near-surface region. So, as both values are still consistent within the error limits, the value resulting from the n-type sample can describe the conductivity of the Si(100)-(2$\times$1) surface state more precisely as $\sigma_{S,Si(100)} = (1.6\,\pm\,0.4) \cdot 10^{-4} \,\mathrm{S}/\square$. 

\section{Conclusion}

\begin{table}[b]
\begin{ruledtabular} 
\centering
\begin{tabular}{ l l }
&\\[-1.5ex]	
Surface reconstruction & Surface conductivity $\sigma_S$\\[1ex]
\hline\\[-1ex]
Si(100)-(2$\times$1) & $(1.6\,\pm\,0.4) \cdot 10^{-4}\,\mathrm{S}/\square$\\ 
Ge(100)-(2$\times$1) & $(3.1\,\pm\,0.6) \cdot 10^{-4}\,\mathrm{S}/\square$\\
Si(111)-(7$\times$7) & $(8.6\,\pm\,1.9) \cdot 10^{-6}\,\mathrm{S}/\square$ \cite{Just}\\
Bi/Si(111)-($\sqrt{3}\times\sqrt{3}$)R$30^{\circ}$ & $(1.4\,\pm\,0.1) \cdot 10^{-4}\,\mathrm{S}/\square$ \cite{Just}\\
Ag/Si(111)-($\sqrt{3}\times\sqrt{3}$)R$30^{\circ}$ & $(3.1\,\pm\,0.4) \cdot 10^{-3}\,\mathrm{S}/\square$ \cite{Luepke}\\[0.5ex]   
\end{tabular}
\caption{Values for the surface conductivity of different reconstructed and passivated surfaces of silicon and germanium.}
\label{table1}
\end{ruledtabular} 
\end{table}

In conclusion, we applied an analytically derived N-layer model for current transport through multiple layers of different conductivity including the calculation of the near-surface band-bending to interpret distance-dependent four-point resistance measurements on semiconductor surfaces.
First, the important role of the space charge region for the current distribution in the sample was discussed and it was shown that already the lowest case of the N-layer model, i.e. the 3-layer model, can describe measured four-point resistance data much better than the often used parallel-circuit model, which completely neglects the space charge region. The derivation of the N-layer model and its usage for multi-probe distance-dependent four-point resistance measurements on surfaces was presented. Finally, the N-layer model was used for describing published distance-dependent four-point measurements on Ge(100) and Si(100) surfaces and values for the conductivities of the surface states of these materials could be determined as summarized in Tab. \ref{table1}. For comparison, values for the surface conductivities of differently reconstructed and passivated Si(111) surfaces are also listed. In total, the presented method is quite generic and can easily be used for many other materials to determine values for the surface conductivity. 

\appendix*

\section{Derivation of the analytical N-layer conductance model} \label{appendix}

\begin{figure}[b!]
\centering
\includegraphics[width=0.315\textwidth]{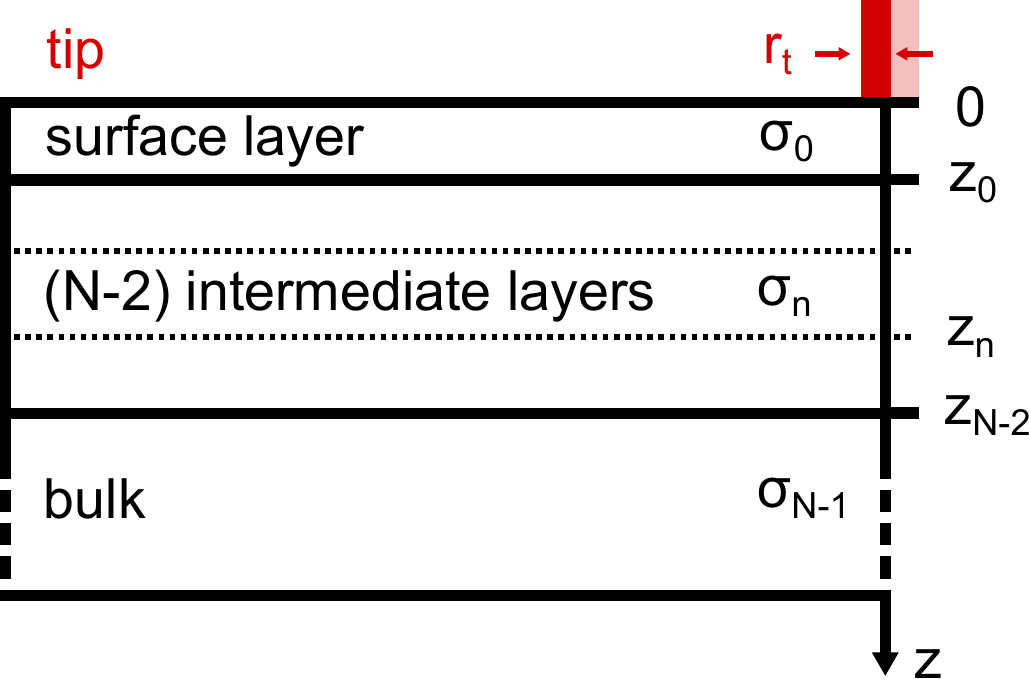}
\caption{(Color online) The N-layer model consists of a layered sample structure with N layers described by the conductivities $\sigma_n$ and the positions of the interfaces $z_n$ ($\mathrm{n} = 1,\ldots,N-2$), respectively. The first layer $0$ and the last layer $N-1$ represent the surface layer and the semi-infinite bulk, respectively. The other layers in between are used to approximate the $z$-dependent conductivity profile of the space charge region. The current $I$ is injected by a cylindrical tip of radius $r_t$ at the origin on the surface layer.}
\label{nlayer}
\end{figure}

The N-layer model uses a structure shown in Fig. \ref{nlayer} to describe the sample properties. It consists of a thin surface layer, multiple intermediate layers for approximating the space charge region and a semi-infinite bulk characterized by their respective conductivities $\sigma_0$, $\sigma_n$ and $\sigma_{N-1}$, and positions of the interfaces $z_0$ and $z_n$ ($\mathrm{n} = 1,\ldots,N-2$). At the surface a current $I$ is injected by a cylindrical tip with radius $r_t$. Due to calculation requirements, the surface layer cannot be two-dimensional, so that a finite thickness of one atomic layer ($3\, \mathrm{\AA}$) is assumed. As $\nabla \cdot \mathbf{j} = 0$ for the current density $\mathbf{j} = \sigma \mathbf{E} = -\sigma\, \nabla \Phi$ inside the sample (excluding the injection point), the electrical potential~$\Phi$ in this region can be determined by solving the Laplace equation  
\begin{align}
	\Delta \Phi & = 0
\end{align}
in cylindrical coordinates. Taking account of the angle-independent polar symmetry for one tip, a solution for the potential in the individual layers is \cite{Jackson} 
\begin{align}
	&\Phi_0(\rho,z) = \int_0^\infty \left[a_0(k) \, e^{kz} + a_1(k) \,e^{-kz}\right] J_0(k\rho) \, \mathrm{d}k \,\mathrm{,} \\[0.5ex] 
	&\Phi_n(\rho,z) = \int_0^\infty \left[a_{2n}(k) \, e^{kz} + a_{2n+1}(k) \,e^{-kz}\right] J_0(k\rho) \, \mathrm{d}k\,\mathrm{,} \\[0.5ex]
	&\Phi_{N-1}(\rho,z) = \int_0^\infty a_{2N-2}(k) \,e^{-kz} \, J_0(k\rho) \, \mathrm{d}k\, \mathrm{,}  
\end{align}
with $J_0$ denoting the Bessel function of the first kind. With the assumption of a uniform current flux beneath the tip contact the boundary conditions are 
\begin{align}
	\sigma_0 \frac{\partial}{\partial z}\Phi_0(\rho,0) & = - j_0\,  H(r_t - \rho)\,\mathrm{,} \label{bound1}\\[0.5ex]
	\Phi_{n-1}(\rho,z_{n-1}) &= \Phi_{n}(\rho,z_{n-1})\,\mathrm{,} \label{bound2}\\[0.5ex]
	\sigma_{n-1} \frac{\partial}{\partial z}\Phi_{n-1}(\rho,z_{n-1}) & = \sigma_{n} \frac{\partial}{\partial z}\Phi_{n}(\rho,z_{n-1})\,\mathrm{,} \label{bound3}\\[0.5ex]
	\Phi_{N-2}(\rho,z_{N-2}) &= \Phi_{N-1}(\rho,z_{N-2})\, \mathrm{,} \label{bound4}\\[0.5ex]
\sigma_{N-2} \frac{\partial}{\partial z}\Phi_{N-2}(\rho,z_{N-2}) & = \sigma_{N-1} \frac{\partial}{\partial z}\Phi_{N-1} (\rho,z_{N-2})\,\mathrm{,} \label{bound5}
\end{align}   
resulting from the current injection (Eq.~\ref{bound1}), as well as from the continuous transitions of the potential (Eq.~\ref{bound2} and Eq.~\ref{bound4}) and the current density (Eq.~\ref{bound3} and Eq.~\ref{bound5}) between the layers. In Eq. \ref{bound1}, the expression $H(r_t - \rho)$ denotes the Heaviside step function. According to the uniform flux condition the injected current density is described by $j_0 = \frac{I}{\pi\,r_t^2}$ assuming a cylindrical tip with a tip radius of $r_t \approx 25\,  \mathrm{nm}$, which seems reasonable for an STM tip. Nevertheless, it turns out that also other values for the tip radius in the range of $5\,\mathrm{nm}$ to $100\,\mathrm{nm}$ do not influence the results of the calculations in a considerable manner. Besides the uniform flux condition \cite{Leong}, several other assumptions for the current density at the injection point have been presented in the literature, i.e. the variable flux condition based on the exact solution for a circular contact on an infinetely thick slab \cite{Schumann} and the Dirac delta current distribution leading to a ring current density \cite{Berkowitz,Wang}. All approaches are used to approximate the exact surface boundary condition of constant potential beneath the probe, which would lead to a more difficult mixed boundary value problem. However, the differences between the three conditions are rather small \cite{Leong2,Berkowitz,Wang}, and especially for small layer thicknesses compared to the radius of the probe contacts, as it applies for the highly conductive surface layer with a thickness of $3\,\mathrm{\AA}$, the uniform flux condition is the best approximation \cite{Leong2}, so that we use this condition for the calculation. 

Based on Eqs.~\ref{bound1} $-$ \ref{bound5}, a matrix equation determining the coefficients $a_0(k),\dotsc, a_{2N-2}(k)$ is derived
\begin{widetext}
\begin{align}
	\begin{pmatrix}
		1 & -1 & 0 & 0 & \ldots & \ldots & \ldots & \ldots & \ldots & \ldots & \ldots & \ldots & 0 \\
		\multicolumn{4}{c}{\multirow{2}{*}{$\mathbf{A}^{0,1}$}} & 0 & 0 & \multicolumn{2}{c}{\multirow{2}{*}{$\ldots$}} & \multicolumn{2}{c}{\multirow{2}{*}{$\ldots$}} & \multicolumn{2}{c}{\multirow{2}{*}{$\ldots$}} & 0 \\
		\multicolumn{4}{c}{} & 0 & 0 & \multicolumn{2}{c}{} & \multicolumn{2}{c}{} & \multicolumn{2}{c}{} & 0 \\
		0 & 0 & \multicolumn{4}{c}{\multirow{2}{*}{$\mathbf{A}^{1,2}$}} & 0 & 0 & \multicolumn{2}{c}{\multirow{2}{*}{$\ldots$}} & \multicolumn{2}{c}{\multirow{2}{*}{$\ldots$}} & 0 \\
		0 & 0 & \multicolumn{4}{c}{} & 0 & 0 & \multicolumn{2}{c}{} & \multicolumn{2}{c}{} & 0 \\
		0 & 0 & 0 & 0 &\multicolumn{4}{c}{\multirow{2}{*}{$\ddots$}} & 0 & 0 & \multicolumn{2}{c}{\multirow{2}{*}{$\ldots$}} & 0 \vphantom{\vdots}\\
		0 & 0 & 0 & 0 &\multicolumn{4}{c}{} & 0 & 0 & \multicolumn{2}{c}{} & 0 \\
		0 & 0 & \multicolumn{2}{c}{\multirow{2}{*}{$\ldots$}} & 0 & 0 &\multicolumn{4}{c}{\multirow{2}{*}{$\mathbf{A}^{n-1,n}$}} & 0 & 0 & 0 \\
		0 & 0 & \multicolumn{2}{c}{} & 0 & 0 &\multicolumn{4}{c}{} & 0 & 0 & 0 \\
		0 & 0 &  \multicolumn{2}{c}{\multirow{2}{*}{$\ldots$}} &  \multicolumn{2}{c}{\multirow{2}{*}{$\ldots$}} & 0 & 0 &\multicolumn{4}{c}{\multirow{2}{*}{$\ddots$}} & 0 \\
		0 & 0 & \multicolumn{2}{c}{} & \multicolumn{2}{c}{} & 0 & 0 &\multicolumn{4}{c}{} & 0 \vphantom{\vdots} \\
		0 & 0 &  \multicolumn{2}{c}{\multirow{2}{*}{$\ldots$}} &  \multicolumn{2}{c}{\multirow{2}{*}{$\ldots$}} &  \multicolumn{2}{c}{\multirow{2}{*}{$\ldots$}} & 0 & 0 & \multicolumn{3}{c}{\multirow{2}{*}{$\mathbf{B}$}} \\
		0 & 0 & \multicolumn{2}{c}{} & \multicolumn{2}{c}{} & \multicolumn{2}{c}{} & 0 & 0 & \multicolumn{3}{c}{} \\ 
	\end{pmatrix}
	\cdot %
	\begin{pmatrix}
		a_0(k)\\
		a_1(k)\\
		a_2(k)\\
		a_3(k)\\
		a_4(k)\\
		\vdots\\
		a_{2n-2}(k)\\
		a_{2n-1}(k)\\
		a_{2n}(k)\\
		a_{2n+1}(k)\\
		\vdots\\
		a_{2N-3}(k)\\
		a_{2N-2}(k)\\
	\end{pmatrix}
	& = %
	\begin{pmatrix}
		I(k,\sigma_0)\\
		0\\
		0\\
		0\\
		0\\
		\vdots\\
		0\\
		0\\
		0\\
		0\\
		\vdots\\
		0\\
		0
	\end{pmatrix}\,\mathrm{,} \label{matrix}
\end{align}
\end{widetext}
with the submatrices
\begin{align}
\mathbf{A}^{n-1,n} =  
	\begin{pmatrix}
	\frac{\sigma_{n-1}}{\sigma_{n}} & -\frac{\sigma_{n-1}}{\sigma_{n}}\, e^{-2 k z_{n-1}} & -1 & e^{-2 k z_{n-1}} \\[1ex]	
	1 & e^{-2 k z_{n-1}} & -1 & -e^{-2 k z_{n-1}}
	\end{pmatrix}
\end{align}
and 
\begin{align}
	\mathbf{B} = 
	\begin{pmatrix}
\frac{\sigma_{N-2}}{\sigma_{N-1}} & -\frac{\sigma_{N-2}}{\sigma_{N-1}}\, e^{-2 k z_{N-2}} & e^{-2 k z_{N-2}} \\[1ex]
1 & e^{-2 k z_{N-2}} & -e^{-2 k z_{N-2}}\\ 
	\end{pmatrix}\,\mathrm{,}
\end{align}
and the expression 
\begin{align}
	I(k,\sigma_0) = - \frac{j_0}{\sigma_0} \, \int_0^{r_{t}} \rho \, J_0(k\rho)\,\mathrm{d}\rho\,\mathrm{.}
\end{align}
This equation can be solved by means of numerical matrix inversion of the $(2N-1) \times (2N-1)$ matrix. As the potential at the surface ($z = 0$) can be expressed by 
\begin{align}
	\Phi_{\mathrm{surf}}(\rho) = \Phi_0(\rho,0) & = \int_0^\infty \left[a_0(k) + a_1(k)\right] J_0(k\rho) \, \mathrm{d}k \,\mathrm{,}
\end{align} 
only the coefficients $a_0(k)$ and $a_1(k)$ are relevant for the calculation. 
Introducing cartesian coordinates with $\mathbf{x} = \begin{pmatrix} x & y \end{pmatrix}^T $ and $\rho = |\mathbf{x} - \mathbf{x_{0}}| = \sqrt{(x-x_{0})^2+(y-y_{0})^2}$ for a tip positioned at $\mathbf{x_0}$, the combined potential on the surface $\Phi_{\mathrm{surf},12}$ for a current source at position $\mathbf{x_{0_1}}$ and a current sink at position $\mathbf{x_{0_2}}$ results by superposition in
\begin{align}
	\Phi_{\mathrm{surf},12}(\mathbf{x}) & = \Phi_{\mathrm{surf},1}(|\mathbf{x} - \mathbf{x_{0_1}}|) -  \Phi_{\mathrm{surf},2}(|\mathbf{x} - \mathbf{x_{0_2}}|) \,\mathrm{.} 
\end{align}
Finally, the four-point resistance $R^{4p}$ measured on the surface is determined by the quotient of the potential difference between the positions $\mathbf{x_{0_3}}$ and $\mathbf{x_{0_4}}$ of the voltage-measuring tips and the current $I$, resulting in 
\begin{align}
	R^{4p} & = \frac{\Phi_{\mathrm{surf},12}(\mathbf{x_{0_3}}) - \Phi_{\mathrm{surf},12}(\mathbf{x_{0_4}})}{I} \nonumber \\[1.5ex]
	       & = \frac{1}{I} \int_0^\infty \left[ a_0(k) + a_1(k) \right] \cdot \left[ J_0(k\, |\mathbf{x_{0_3}} - \mathbf{x_{0_1}}|) \right. \nonumber \\[1ex]
		&\hspace{4.75ex}\left. - J_0(k\, |\mathbf{x_{0_3}} - \mathbf{x_{0_2}}|) %
	\hspace{0ex} - J_0(k\, |\mathbf{x_{0_4}} - \mathbf{x_{0_1}}|) \right. \nonumber \\[1ex]
		&\hspace{4.75ex} \left. + J_0(k\, |\mathbf{x_{0_4}} - \mathbf{x_{0_2}}|)  \right] \, \mathrm{d}k\, \mathrm{.}  \label{r4p}
\end{align}
For the linear probe configuration with equidistant spacing $s$ between the four tips Eq. \ref{r4p} simplifies to 
\begin{align}
	R^{4p}(s)  & = \frac{2}{I} \int_0^\infty \left[ a_0(k) + a_1(k) \right] \cdot \left[ J_0(k s) - J_0(2 k s) \right] \, \mathrm{d}k \, \mathrm{.} \label{r4p-linear}
\end{align}

The integral over the Bessel functions in Eq. \ref{r4p-linear} can be evaluated numerically and the result can be fitted to the four-point measurement data. However, the conductivities $\sigma_n$ and interface positions $z_n$ of all N layers are far too many free parameters for being determined by a single fit. Therefore, the depth-dependent conductivity profile $\sigma(z)$ of the space charge region has to be calculated before, based on the solution of Poisson's equation using basic material parameters like the Fermi level pinning of the surface states, the band-gap, the effective masses, the mobilities and the bulk doping concentration and type \cite{Lueth}. The approximation of this profile by a step function of $(N-2)$ - steps then determines the values for $\sigma_1,\hdots,\sigma_{N-2}$ and $z_1,\hdots,z_{N-2}$, which are used as input for the N-layer model. The thickness of the surface layer $z_0$ determines the vertical extension of the surface states and can be approximated by the thickness of one atomic layer.    
The value for the bulk conductivity $\sigma_{N-1}$ can be determined by macroscopic resistivity measurements and should be in agreement with the nominal doping concentration. Finally, only one free parameter remains to be determined by a fit to measurement data, which is the surface conductivity $\sigma_0$.    

\bibliography{./nlayer-paper_240815}

\end{document}